\documentclass[preprint,journal]{vgtc}            % preprint (journal style)

\onlineid{1895}

\vgtccategory{Research}

\vgtcpapertype{theory/model}

\title{MENA: Multimodal Epistemic Network Analysis for Visualizing Competencies and Emotions}

\author{%
  Behdokht Kiafar,
  Pavan Uttej Ravva,
  Asif Ahmmed Joy,
  Salam Daher, and
  Roghayeh Leila Barmaki
}

\abstract{%
 The need to improve geriatric care quality presents a challenge that requires insights from stakeholders. While simulated trainings can boost competencies, extracting meaningful insights from these practices to enhance simulation effectiveness remains a challenge. In this study, we introduce Multimodal Epistemic Network Analysis (MENA), a novel framework for analyzing caregiver attitudes and emotions in an Augmented Reality setting and exploring how the awareness of a virtual geriatric patient (VGP) impacts these aspects. MENA enhances the capabilities of Epistemic Network Analysis by detecting positive emotions, enabling visualization and analysis of complex relationships between caregiving competencies and emotions in dynamic caregiving practices. The framework provides visual representations that demonstrate how participants provided more supportive care and engaged more effectively in person-centered caregiving with aware VGP.
This method could be applicable in any setting that depends on dynamic interpersonal interactions, as it visualizes connections between key elements using network graphs and enables the direct comparison of multiple networks, thereby broadening its implications across various fields. {The code and setup to reproduce the experiments are publicly available \href{https://anonymous.4open.science/r/MENA-E668/README.md}{\color{blue}here}, and data is available upon request.}
}


\teaser{
  \includegraphics[width= 1\textwidth]{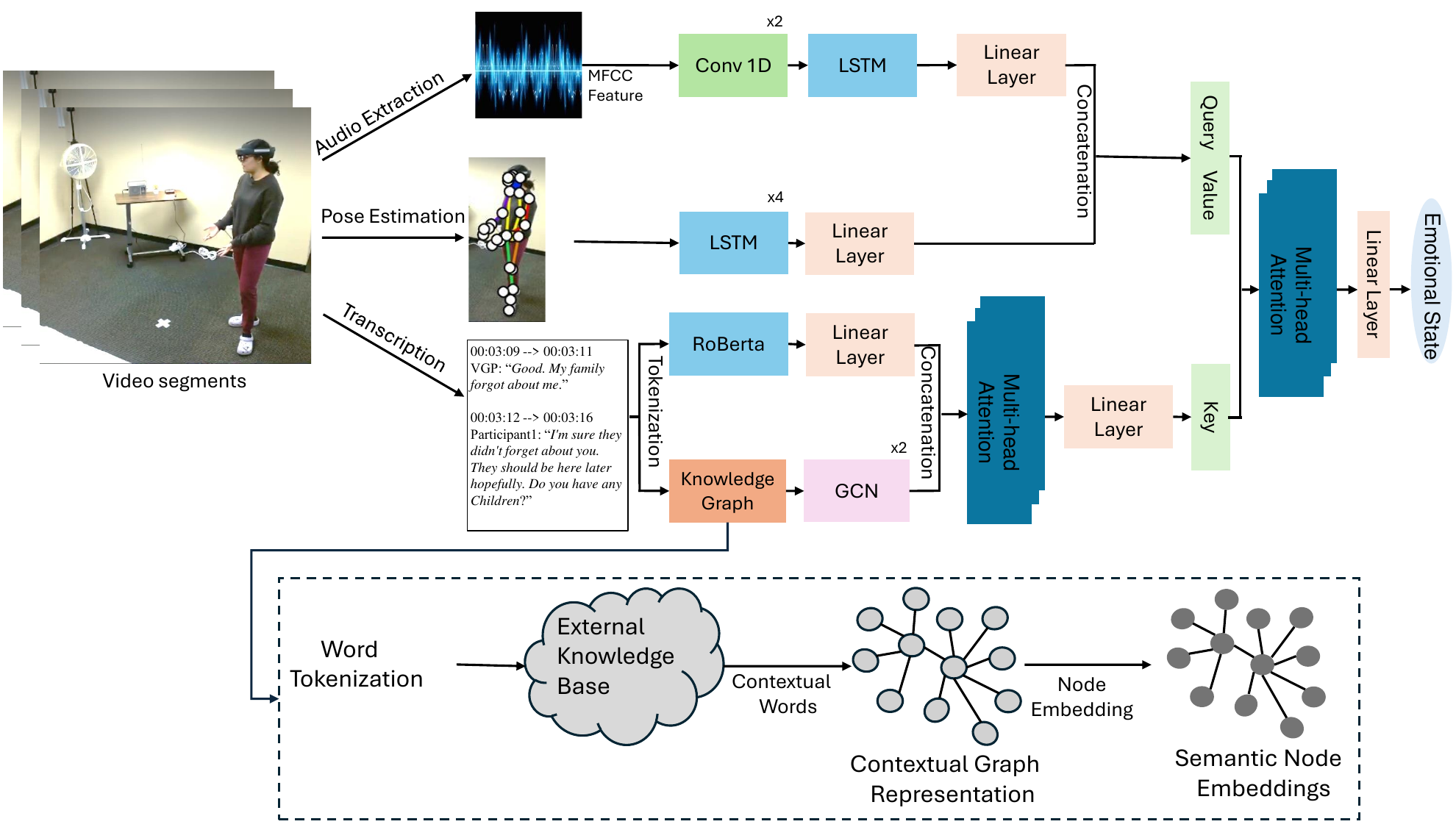}
  \caption{The architecture of the proposed multimodal Emotional State Classifier, consisting of four main components: Audio Extraction, Pose Estimation, Text Feature integrated with Knowledge Graph, and Fusion Network including a classification head. This setup determines whether the participant exhibited a positive emotion, such as a supportive and uplifting approach, during that segment of the video, as evidenced by their audio, gestures, and verbal communication.
  }
  \label{fig:teaser}
}

\graphicspath{{figs/}{figures/}{pictures/}{images/}{./}} % where to search for the images

\usepackage{tabu}                      % only used for the table example
\usepackage{booktabs}                  % only used for the table example
\usepackage{lipsum}                    % used to generate placeholder text
\usepackage{mwe}                       % used to generate placeholder figures

\usepackage{mathptmx}                  % use matching math font

\usepackage{float}
\usepackage{comment}

\usepackage{amsmath}

\begin{document}

%%%%%%%%%%%%%%%%%%%%%%%%%%%%%%%%%%%%%%%%%%%%%%%%%%%%%%%%%%%%%%%%
%%%%%%%%%%%%%%%%%%%%%% START OF THE PAPER %%%%%%%%%%%%%%%%%%%%%%
%%%%%%%%%%%%%%%%%%%%%%%%%%%%%%%%%%%%%%%%%%%%%%%%%%%%%%%%%%%%%%%%

\firstsection{Introduction}

\maketitle

By 2030, an estimated 21.6\% of the U.S. population will exceed 65 years old \cite{acl_profile_2023}. This trend underscores the rising demand for caregiving professionals, intensified by the high turnover rate and persistent shortage of caregivers at present moment \cite{rollison2023evaluation}. A potential factor behind the elevated turnover rate is insufficient training, which is particularly concerning given the complex nature of interactions with older adults \cite{lim2021characteristics}.

In this context, immersive solutions such as Augmented Reality (AR), and Virtual Reality (VR) technologies provide risk-free tools in healthcare simulation, enhancing training experiences \cite{kanschik2023virtual}. Simulations can offer immersive and interactive environments that can potentially equip caregivers with the necessary skills to meet the complex needs of the elderly.
Recognizing the limitations of traditional training methods in adequately preparing caregivers, AR and VR technologies have been developed to create a more dynamic and engaging learning environment by integrating virtual humans that can simulate real-world interactions and scenarios.

As AR and VR technologies continue to provide new opportunities for caregiver training experiences, it is essential to evaluate the outcomes of these simulations to develop the most effective practices. Caregiving process is a dynamic and multimodal phenomenon that involves interactive, cognitive, behavioral, and emotional aspects of communication that can affect the quality of care provided. Thus, a holistic multimodal analysis would help to understand these interconnected interactions.
However, integrating various data types poses a challenge, as it necessitates processing, annotating, and analyzing the raw data to yield meaningful and interpretable results for researchers and practitioners \cite{cukurova2020promise}. Nevertheless, there has been notable progress in combining different data sources to enrich the understanding of such training systems. For instance, a recent study by Vatral et al. \cite{vatral2022using} developed a method to improve traditional qualitative analysis by integrating multimodal learning analytics, using data collected during simulation-based training for nurses within a mixed reality manikin-based environment. In another research effort \cite{fernandez2022effectiveness}, a multimodal intervention has been shown to enhance the attitudes of nursing students toward older adults. This was assessed using psychometric instruments to evaluate the effectiveness of simulation-based training.

However, these studies do not fully capture how participants connect with the patient at a high level, nor do they consider the emotional dimensions involved. Generating such a picture to understand these aspects comprehensively is crucial for the effectiveness of caregiving simulation applications.
In addition, previous research has identified a significant gap in empathy and emotional intelligence between formal training and the actual experiences of caregivers \cite{kiafar2024analyzing, lee2021effectiveness}. These studies underscore the need for training scenarios that are more emotionally engaging in simulations of geriatric care.

To address this gap, this study introduces a novel approach, Multimodal Epistemic Network Analysis (MENA), to analyze the interplay of contextual dialogue data and emotional states of participants interacting with a virtual geriatric patient (VGP) in an AR environment.
Epistemic Network Analysis (ENA) within Quantitative Ethnography (QE) \cite{shaffer2017quantitative} is recognized in behavioral analytics, providing interpretable insights into how behaviors are interconnected \cite{elmoazen2022systematic}. This method maps the relationships between events to identify which behaviors commonly occur together.
We have enhanced this method by developing a multimodal Emotional State Classifier (ESC) and incorporating the emotional states of participants into the network analysis, providing a more nuanced understanding of the affective interactions.

To evaluate our proposed approach, we applied it within a virtual geriatric simulation, assessing caregiver performance. We designed two modes of interaction: an \textit{unaware} mode, where the VGP was not conscious of its surroundings, and an \textit{aware} mode, where the VGP demonstrated awareness of environmental changes and previous conversations. We compared these modes using a within-subjects pilot study involving 20 participants.
Visual representations and statistical measurements of MENA allow us to examine the temporal relationships between nursing competencies and expressed emotions across participants, gaining a better understanding of the optimal conditions for simulation training.
Our contributions are as follows:

\begin{itemize}
  \item Developing a multimodal Emotional State Classifier that enhances the accuracy of capturing positive emotions compared to single-modality approaches.
  \item Proposing a Multimodal Epistemic Network Analysis framework
that leverages visual tools to examine and quantify nursing competencies and emotional responses of caregivers interacting with a VGP.
  %enhancing our understanding of their interactions within the simulation.
  \item Results of a pilot user study (n = 20), demonstrating how behaviors of participants change in response to different simulation conditions across two distinct interaction modes with the VGP.
  %providing empirical insights into the impact of the VGP's awareness levels on competencies and emotional responses of participants
\end{itemize}

\section{Background and Related Work}

\subsection{Caregiver Training through Virtual Simulation}

Technology-based programs provide experiential settings where nursing students can learn and practice in simulation environments \cite{koukourikos2021simulation}. These systems offer a safe and controlled environment that can improve the caregiving abilities and confidence of trainees \cite{khalil2023effect}. This strategy bridges the gap between theoretical knowledge and practical application, boosting the competencies of nursing students in clinical settings \cite{azizi2022comparison}.
%Early qualitative work on virtual simulation training found that these programs can better prepare and equip nursing students with necessary competencies when students practice with interactive virtual environments \cite{lange2020learning}.
To further enhance these simulation experiences, the integration of AR in nursing education has shown significant potential \cite{wuller2019scoping}. By providing immersive learning experiences, this technology prepares students for real-world clinical practice, allowing them to engage in realistic scenarios and develop the critical skills necessary for patient care \cite{quqandi2023augmented}.

Nakazawa et al. \cite{nakazawa2023augmented} explored the use of AR for affective training and found that it enhanced communication skills and empathy among caregivers of people with dementia.
Yoo et al. \cite{yoo2024adoption} discussed how using AR in critical care training enhanced self-efficacy, familiarity, and confidence, while also reducing anxiety.
Uymaz et al. \cite{uymaz2022assessing} examined the acceptance of AR technology among nursing students and found a high intention to use AR for self-learning. This study also noted the potential of AR to complement traditional learning methods by offering interactive and engaging educational experiences.

While this technological advancement provides promising avenues for enhancing nursing education, it is essential to consider the methodologies used to assess the effectiveness of these systems.
Across these investigations, evaluations predominantly use subjective tools such as the System Usability Scale \cite{lewis2018system}, the Technology Acceptance Model \cite{marangunic2015technology}, and the Interpersonal Reactivity Index \cite{cronin2018interpersonal}, or rely on qualitative assessments through semi-structured interviews, which may introduce biases.

Recognizing these potential biases and limitations, the adoption of Quantitative Ethnography could provide a more robust framework for analyzing the intricate dynamics of practical skills and competency development, offering a quantitative lens to evaluate the impact and effectiveness of educational technologies \cite{shaffer2017quantitative}.

\subsection{Quantitative Ethnography and ENA in Nursing Education}
Quantitative Ethnography is a methodological framework that integrates the detailed, contextual insights of ethnographic (qualitative, case-based) methods with the pattern-detection capabilities of quantitative (statistical, data-driven) approaches to study human behavior \cite{kaliisa2021scoping}. While quantitative methods excel at identifying patterns in numerical data, they can miss nuanced details. Conversely, qualitative techniques offer in-depth insights but can be subject to bias and may not always be generalizable. By merging these two approaches, QE enhances our ability to understand and model complex human behaviors.
Within Quantitative Ethnography framework, the ENA approach supports analysis by visually representing the connections within qualitative data, along with statistical measures, providing a structured method to uncover complex interactions.

Weiler et al. \cite{weiler2022quantifying} applied ENA to study dementia caregiver work systems, using semi-structured interviews to identify work system components and create configural diagrams. Their analysis revealed five key interactions and compared work system configurations of caregivers providing care at home and away from home. Shah et al. \cite{shah2022quality} employed ENA to examine how nursing students develop competencies like teamwork, collaboration, and safety by analyzing faculty and student conversations during virtual reality simulations and debriefing sessions. Kiafar et al. \cite{kiafar2024analyzing} demonstrated a discrepancy between training and the actual practice of caregivers by visualizing the connections between different aspects of caregiving. This revealed a misalignment between the preparation of nursing assistants and their real-life clinical expectations, which was uncovered in the interview data using ENA.

While insightful, these studies focused on a single modality and overlooked the potential benefits of simultaneously exploring various types of data. Few studies have considered employing a multimodal approach in QE to enrich the analysis. 
For example, Shum et al \cite{buckingham2019multimodal} introduced the Multimodal Matrix as data modeling method within QE, combining theoretical concepts of teamwork with contextual observations from a nursing simulation lab setting. In subsequent studies, \cite{echeverria2019towards, martinez2023lessons}, various data types, including video and audio recordings, have been integrated to map higher-level concepts with low-level sensor data to provide a comprehensive view of student interactions and behaviors, using QE to interpret them and enhance the overall analysis.
To the best of our knowledge, no systematic study has investigated behavioral dynamics in a virtual caregiving setting using a multimodal approach. This gap underscores the need for a foundational basis for evaluating such interactions comprehensively. Bridging this gap, we extend our analysis beyond traditional conversational data by introducing emotional states as a pivotal modality in our QE framework, as caregivers need to effectively manage and express their emotions to maintain the quality of care they provide \cite{monin2009willingness}. 

\subsection{Emotion Classification and Prediction Models}

Numerous studies have explored Emotional State Prediction (ESP) in conversations to better understand the well-being of individuals, emphasizing the importance of context \cite{fu2023emotion}. However, emotions are conveyed through different modalities, and integrating these modalities offers a more comprehensive understanding of emotional expression. Recognizing this complexity, researchers have started to explore multi-modal approaches, focusing on text and audio data  \cite{hung2023beyond, hema2023emotional}. In addition, with the growing availability of open-source datasets \cite{zadeh2018multi, livingstone2018ryerson}, multimodal data analysis has gained popularity, combining visual data with audio and text.

Along with these advancements in multimodal analysis, the techniques used for data extraction and analysis have also progressed. Traditionally, researchers manually extracted Mel Frequency Cepstral Coefficient features (MFCC) from audio to train models like k-Nearest Neighbor and Support Vector Machines for emotion classification \cite{likitha2017speech}. However, with the rising demand for deep learning, Convolutional Neural Networks (CNNs) have become more popular, leading researchers to adopt them for various tasks \cite{patni2021speech}. Meanwhile, the introduction of Transformers has significantly impacted text-based ESP by enabling models to process large amounts of information \cite{acheampong2021transformer}. Models like BERT \cite{devlin2018bert} and GPT leverage pre-trained contextual embeddings and transfer learning, capturing nuanced emotional expressions more effectively than traditional methods. DialogueGCN \cite{ghosal2019dialoguegcn} uses a graph CNN to understand the relationships between words, showcasing the effectiveness of treating words as a graph structure.
Yahui Fu et al. \cite{fu2022context} proposed a multimodal framework that integrates audio, text, and a knowledge graph, demonstrating significant improvements in ESP. Vishal et al. \cite{chudasama2022m2fnet} developed a multimodal framework that integrates audio, text, and visual data using multi-head fusion layers.

Although these studies have mainly focused on text and audio, they do not capture a wider range of emotional expressions across different modalities. For instance, body gestures are crucial non-verbal cues that can enhance emotional recognition \cite{kleinsmith2012affective}. In this work, we integrate 3D-skeleton pose data, text, and audio, along with a knowledge graph to predict emotions. Using skeleton pose data enhances privacy by not exposing individual identities. Additionally, the knowledge graph offers deeper insights into the conversation by incorporating common-sense concepts from the ConceptNet corpus \cite{speer2017conceptnet}.

\section{Study Framework}
The research questions (RQs) guiding our study are as follows:

\begin{enumerate}[label=\textbf{RQ\arabic*:}, leftmargin= 20mm]
\item How effectively can we classify the emotional states of participants during interactions with the VGP using our proposed framework?
\item How can connections between core nursing competencies and socio-emotional responses be visually represented as participants interact with the VGP?
\item To what extent does the structure of these connections vary among participants when interacting with an \textit{aware} versus \textit{unaware} VGP?
%\item How does prior experience with VR technology influence these connection patterns?
\end{enumerate}

\noindent In the following sections, we describe the participants, apparatus, study procedure, and study design in detail.

\subsection{Participants and System Setup}

IRB approval was obtained, and flyers were distributed at the affiliated institution's college of nursing to recruit participants. The inclusion criteria for participation required an educational background in nursing and normal or corrected-to-normal vision. Interested individuals reached out, and sessions were scheduled accordingly. Demographic details of the participants are provided in Table \ref{tab:participants}.

\begin{table}
\caption{Participant background and characteristics ($n=20$).}
\label{tab:participants}
\renewcommand{\arraystretch}{0.8}  % Adjust the factor to decrease the space between rows
\begin{tabular}{p{0.44\columnwidth}p{0.14\columnwidth}p{0.27\columnwidth}}
\hline
Characteristics  & Value  & Mean \\
\hline
Age   & [18 -- 68] & 28.1\,$\pm$\,15.9 \\
Experience with elderly care   & [0 -- 30] & 4.5\,$\pm$\,8.4 \\
Gender                    &    &    \\
\hspace*{0.5cm} Male      & 2 & (10\%) \\
\hspace*{0.5cm} Female    & 18 & (90\%) \\
Education                &   &      \\
\hspace*{0.5cm} Bachelor's program  & 18 & (90\%) \\
\hspace*{0.5cm} University faculty & 2 & (10\%) \\
VR experience
           &    &     \\
\hspace*{0.5cm}Never used   & 10 & (50\%) \\
\hspace*{0.5cm}Used before & 10 & (50\%) \\

\hline
\end{tabular}
\end{table}

\begin{figure*}[h]
    \centering
    \begin{minipage}{0.8\textwidth}
        \includegraphics[width=\textwidth]{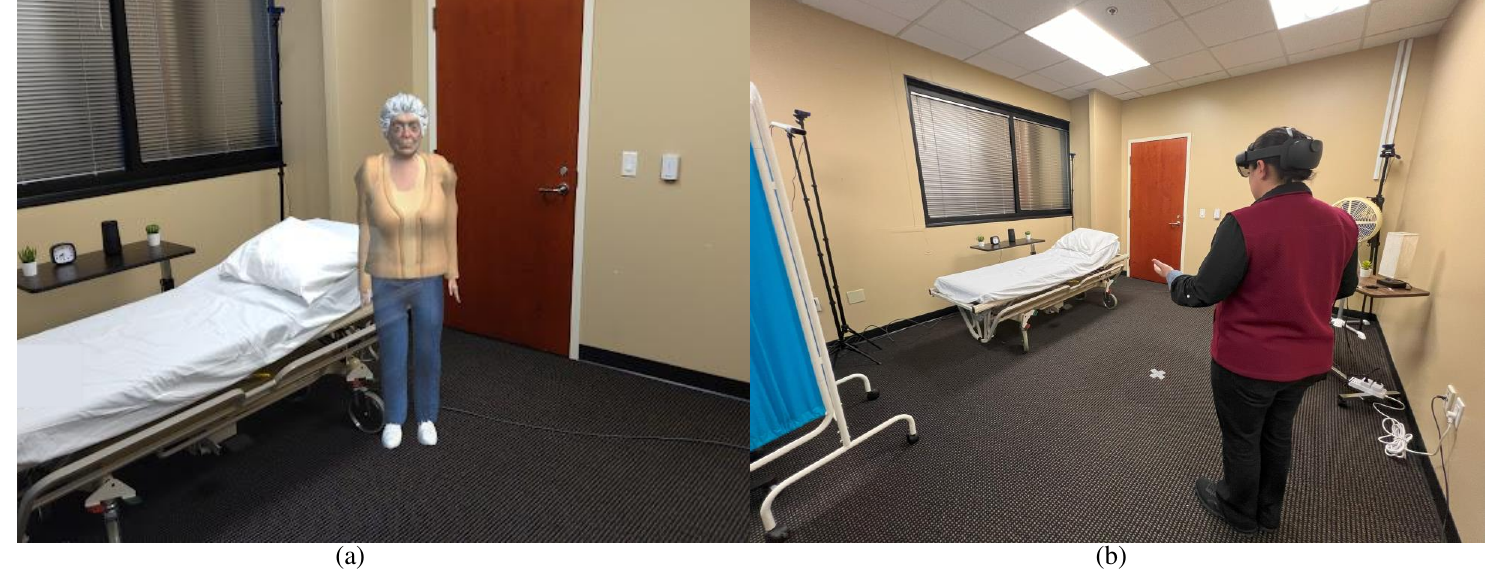}
        \caption{Augmented Reality setup and environment for caregiving practice: (a) the 3D representation of a virtual geriatric patient in a nursing home setting. (b) interaction between a participant and the VGP.}
        \label{fig:Apparatus}
    \end{minipage}
\end{figure*}

% We used the Microsoft HoloLens 2 mixed reality headset to display the VGP, which was developed using the Unity game engine (version 2019.4.34f 1). The 3D model was customized using Autodesk Maya, with 3D scans from 3dsk and Turbosquid as references, reflecting the intricacies of real human anatomy and skin texture. This virtual avatar, animated as an older woman, presents gestures and facial expressions, further enhancing user engagement.
We used a Human-in-the-Loop approach to manage the responses of the VGP in the user study to ensure swift and coherent replies. The VGP response control system includes a Graphical User Interface (GUI), a response control database, and a response data transmission mechanism. 
Additionally, we developed a custom Unity plugin to enable the simultaneous operation of lip synchronization, facial expressions, audio playback, and body gestures of the VGP.
% This process is facilitated by a real-time, bidirectional TCP server-client communication implemented through socket programming in Unity. Using SALSA Lipsync Suite Pro (v2) module, a Unity plugin was developed to dynamically activate lip synchronization and facial expressions on the VGP during user interactions.
Fig. \ref{fig:Apparatus} illustrates a representation of this VGP and setting configuration. We also considered an awareness feature in the framework, enabling the VGP to recognize environmental changes, previous and ongoing conversations, and surroundings. This feature was utilized in the study experiment with further details in the next section.

\subsection{Study Procedure}

First, participants completed consent and prequestionnaire forms soliciting their demographics and background experiences. 
They were then introduced to the AR headset to ensure comfort and familiarity with the technology. The study commenced as participants entered a residence akin to a nursing home setting, where they interacted with a VGP and engaged in typical caregiver-patient conversations.
During this initial phase, the VGP operated in an \textit{unaware} mode, oblivious to environmental changes and prior dialogues. After completing it, participants filled out a mid-questionnaire during a short break. Following this, participants re-entered the residence. In the second phase, the VGP was in \textit{aware} mode, demonstrating awareness of her surroundings and previous interactions. She remembered personal details about the participants and could control IoT devices, like lamp and radio, to reflect this awareness. After this phase, participants took another break to complete a post-questionnaire. Moreover, during one of the conditions, participants played a 20-questions game with the VGP, a common activity between patients and caregivers, to foster a more natural interaction.

\subsection{Study Design}

Our study was designed as a $2 \times 1$ within-subject experiment, with VGP awareness as the independent variable. The dependent variables were the MENA network graphs, which illustrated the relationships between the emotional states of participants and their demonstrated nursing competencies.
Audio and video of the participants were recorded throughout the study for use in our analytical frameworks.
%More details are provided in the next section.

\section{ Methodology} 

This section describes the theoretical foundation of our proposed analytic tool: Multimodal Epistemic Network Analysis. We begin by explaining the data preparation process, which is critical for integration into our analytical technique. Following this, we describe ENA, and then present our framework for emotional state classification. Finally, we introduce MENA, our advanced method that expands ENA by incorporating new construct, including the emotional states of participants.
\subsection{Data Preperation}

For an in-depth analysis, different modalities were extracted and processed from the recorded data. Initially, the video data was segmented by turns of talk between the participant and the VGP. This data was then fed into different algorithms to extract audio signals, 3D skeleton poses, and transcriptions of the conversations, as described in the following sections:

\subsubsection{Audio Extraction}
To extract the audio from each video segment, we used the ffmpeg library in Python. This method enables the extraction of the audio stream from the video, saving the audio from each segment as an individual file. Subsequently, the extracted audio files were used to generate transcription data. 

\subsubsection{3D Human Body Pose}
The 3D skeletal pose data, which provides information about the coordinates of human joints in three-dimensional space, was extracted using the MediaPipe framework \cite{lugaresi2019mediapipe}. 
This framework is built by using nodes known as calculators, which help in feature extraction and image transformations. These calculators enable accurate detection of joint positions and provide the 3D coordinates for each joint across the body.

For each segment of the video, the extracted 3D human body pose data was stored in a .csv file format. Each row in the file corresponds to a specific frame of the video, while the columns represent the 3D coordinates of the various joints. To normalize this data for training, the number of frames across all video segments was standardized to ensure consistency.
% If a segment contained fewer frames than the calculated average, rows were duplicated to match the required length. Conversely, if a segment contained more frames than the average, the excess rows were truncated. 

\subsubsection{Transcriptions of Conversation}
The audio data was utilized to extract the corresponding textual information through an automated transcription process. For this task, AWS Transcribe was employed, and each audio segment was processed to generate a detailed transcription.

% Having prepared the data, the analysis procedure is described in the following sections.

\subsection{Epistemic Network Analysis}
ENA is an analytical technique originally designed to model networks of discourse that emphasize the significance of relationships between ideas. It provides a mathematical method to represent individuals' epistemic frames as comprehensive networks, including their epistemologies and beliefs, in an interpretable format. It has been applied to model discourse networks \cite{goldfarb2024modeling}, trajectories of cognitive and social networks \cite{brohinsky2021trajectories}, social gaze coordination \cite{andrist2015look}, and healthcare studies \cite{zorgHo2023epistemic}.

\subsection{Emotional State Classifier}
In this section, we explain the design of our proposed Emotional State Classifier, which captures instances when the participant expresses positive emotions.
%communicates in a caring and supportive manner, with the aim of creating a positive effect.
To assess the reliability of the labeling process, two of the researchers independently labeled a random sample of 40 segments. They assigned a ``1” to any speech turn that the participant demonstrated positivity, warmth, and empathy, making the VGP feel more hopeful or inspired, and a ``0” otherwise.
The inter-rater reliability (IRR) via Cohen’s kappa \cite{mchugh2012interrater}, was $0.81$, indicating a strong agreement between the two raters, and the percentage of agreement was $92.5\%$.
Based on this high level of agreement, one of the researchers proceeded to label the remaining video data.
The ESC consists of four different modules, as depicted in Fig. \ref{fig:teaser}:

\subsubsection{Audio Feature Extraction}

The audio data was converted into a numerical format for training by transforming the audio signals into Mel Frequency Cepstral Coefficients, which capture audio features aligned with human auditory perception. This process involved dividing the audio signal into frames, calculating the energy spectrum and Fourier transform, and mapping the results to Mel-frequency bins. Then a Discrete Cosine Transform was applied to extract 40 MFCC coefficients \cite{patni2021speech}. The extracted MFCC data was then fed into two one-dimensional CNN layers, followed by a max pooling layer, and subsequently processed through two LSTM layers to learn the sequential relationships within the data. The final features were then sent to the fusion network, as explained in Section \ref{sec:Fusion Network}.

\subsubsection{Human Pose Estimation}
Previous work has predominantly focused on extracting visual features from facial expressions; however, body language also provides valuable emotional insights \cite{de2006towards}, and can address privacy concerns associated with facial data. The extracted skeleton data was fed into an LSTM network designed to handle 3D coordinates, consisting of four hidden layers with 64 neurons each. This configuration effectively captures complex patterns while balancing performance and computational efficiency. Our experiments showed that the model with four hidden layers outperformed others in terms of both performance and computational load. The output of the LSTM layers is subsequently processed through a linear layer and a fusion network, with more details provided in section \ref{sec:Fusion Network}.

\subsubsection{Knowledge Graph and Text Feature Integration} 

%In this work, we utilized ConceptNet \cite{speer2017conceptnet}, which provides a rich base of commonsense knowledge. Concept-Net is multilingual graph connections between words and phrases with labeled weights which gives deeper meaning of words.

In this work, we utilized ConceptNet \cite{speer2017conceptnet}, a rich repository of common-sense knowledge available in multiple languages. ConceptNet is a graph that links words and phrases with labeled weights, providing a deeper understanding of their meanings. To extract common-sense information, the text data of utterances is first converted into tokens, which are then processed through the knowledge graph (KG). Each related word and its corresponding token are treated as nodes, interconnected by edges in the graph. Nodes are embedded using ConceptNet Numberbatch Embeddings to form a graph representation. To analyze the relationships between nodes in the graph, we used a Graph Convolutional Network (GCN) \cite{yao2019graph}, which learns the embedding relations among all nodes.

Furthermore, for a deeper understanding of the utterance context, we integrated the KG with the RoBERTa model \cite{liu2019roberta}, which captures nuanced language patterns and contextual information. The RoBERTa model was fine-tuned using textual data extracted from interactions between caregivers and the VGP. The embeddings generated by RoBERTa were then combined with those obtained from the knowledge graph, creating a more comprehensive representation of the utterances. Finally, the combined features were fed into the fusion layer, detailed in the following section.

\subsubsection{Fusion Network and Classification Head}
\label{sec:Fusion Network}
The Fusion Attention Layer integrates features from multiple modalities by concatenating the final features from the audio feature extractor, the human pose extractor, and KG's features combined with text features. The concatenated features vector of KG's and text is passed through the linear layer, and the resulting output is used as Key (K) in the attention mechanism. Similarly, concatenated features from of audio and pose extractor are used as Query (Q) and Value (V). We used a four multi-head attention mechanism to enable effective integration and alignment across modalities. The outputs of this attention layers form a unified multimodal representation, which is then passed through a classification head that consists of a linear layer for predicting the emotional class.

\subsection{Multimodal Epistemic Network Analysis}
Building on ENA, our proposed method, MENA, extends this approach by incorporating an additional data modality: the emotional states of participants during the interaction with VGP.
MENA specifically examines whether participants convey empathetic support and validate patient emotions throughout compassionate interactions.
This enhancement allows for a more detailed and interpretable model of the connections within an epistemic frame. In the following, this methodology is explained step by step, with an excerpt of an example taken from real coded data to illustrate its application.

\subsubsection{Data Formatting and Coding}
Correctly formatting and coding of multimodal data is essential for employing MENA to model the connections among the data. In the context of ENA, codes are used to categorize specific aspects of the data being analyzed, with each code representing a distinct concept \cite{shaffer2016tutorial}.

Our initial coding scheme was developed based on ``THE ESSENTIALS: CORE COMPETENCIES FOR PROFESSIONAL NURSING EDUCATION \cite{american2021essentials}”, which outlines ten domains of nursing competency, along with sub-competencies. Our analysis focused on participants' practice of competencies within domains 1 (\textit{Knowledge for Nursing Practice}), 2 (\textit{Person-Centered Care}), 5 (\textit{Quality and Safety}), and 9 (\textit{Professionalism}). In addition, we used \textit{Patient Vulnerability Disclosure} to capture instances where VGP requires more compassionate care. These specific domains were selected because they are particularly applicable to the context of our study and the nature of interactions between participants and VGP. 
 We further enhanced our analysis by integrating \textit{Positive Emotion}, derived from the output of the Emotional State Classifier, as an advanced code to enhance our analysis.
Table \ref{tab:coding-scheme} displays the schematic version of the codebook for our study, including the codes and their corresponding definitions.

\begin{table*}
\centering
\caption{Coding scheme for analyzing caregiver competencies in an AR geriatric simulation, adapted from \cite{american2021essentials}.}
\label{tab:coding-scheme}
\begin{tabular}{lp{0.6\textwidth}}
    \toprule
    Code & Definition \\
    \midrule
    Knowledge for Nursing Practice & Integration and application of comprehensive nursing knowledge along with interdisciplinary insights. \\[0.9ex]  % Adjust the spacing here
    Person-Centered Care & Providing individualized, evidence-based care that is respectful, compassionate, and informed by a deep understanding of the individual's life experiences. \\[0.5ex]  % Adjust the spacing here
    Quality and Safety & Employment of established principles of safety and improvement science to enhance care quality and minimize harm to patients and providers. \\[0.9ex]  % Adjust the spacing here
    Professionalism & Continuous personal and professional growth, emphasizing ethical comportment, social responsibility, and advocacy for social justice. \\[0.9ex]  % Adjust the spacing here
    Patient Vulnerability Disclosure & Occasions when the VGP displays feelings that call for increased compassionate care. \\[0.9ex]  % Adjust the spacing here
    Positive Emotion & Expressing positive emotions or attitudes like empathy and supportiveness to enhance mood by acknowledging and responding to VGP's feelings. \\  % Adjust the spacing here
    \bottomrule
\end{tabular}

\end{table*}

\begin{figure*}[h]
    \centering     \includegraphics[width=0.9\textwidth]{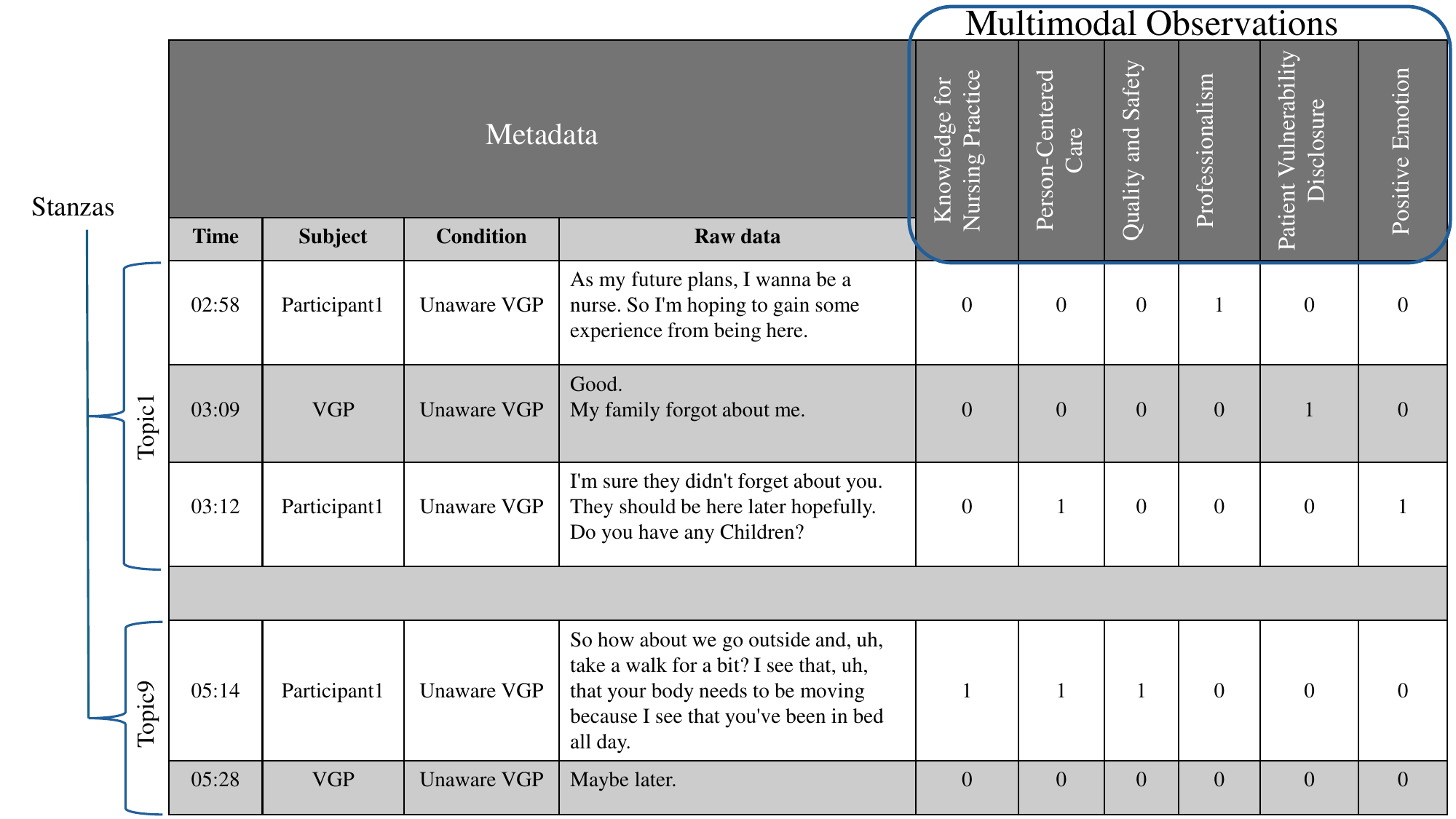}
    \caption{Excerpt of coded data with a schematic design of the Multimodal Matrix used in MENA.}
    \label{fig:Coded_Data}  
\end{figure*}

To understand the coding process, it is useful to examine an example from the actual coded data of participants during the study (see Fig. \ref{fig:Coded_Data}).
This sample includes the transcriptions noted in the column titled ``Raw data”. Each line corresponds to an utterance from either the participant or the VGP, as indicated in the ``Subject” column. The ``Condition” column shows whether the interaction is in the \textit{aware} or \textit{unaware} mode. The six rightmost columns contain codes with binary values that indicate the presence ($1$) or absence ($0$) of each code within the utterances.
Additionally, conversations are organized into ``Stanzas” based on topics, with each distinct topic considered as a separate stanza. The rationale behind this structure is explained in further detail in section \ref{Constructing adjacency matrices}.

%The five rightmost columns, referred to as the code columns, display the nodes of the network model, allowing for the visualization of code co-occurrence within the same line of data.

%\vspace{-2mm}

The transcripts were segmented into lines corresponding to turns of talk, for a total of 3,710 lines, and coded using the six codes shown in Table \ref{tab:coding-scheme}.
80 lines of data were coded for the first five codes using the automated classifier nCoderR package \cite{ncodeR} and by two team researchers. The kappa and rho values between the computer and the first human rater were $0.76$ and $0.03$, respectively. For the computer and the second human rater, the kappa was $0.83$, and the rho was $0.04$. Given the high levels of agreement across all rater pairs, it was concluded that the automated classifier was valid and reliable for coding the remaining data, minimizing human intervention.
Additionally, the \textit{Positive Emotion} code was derived from the Emotional State Classifier.
Our units of analysis were the individual participants across different scenarios.

\subsubsection{Constructing Adjacency Matrices to Demonstrate the Co-occurrence of Codes in Each Stanza}
\label{Constructing adjacency matrices}
\begin{figure*}
    \centering     \includegraphics[width=0.8\textwidth]{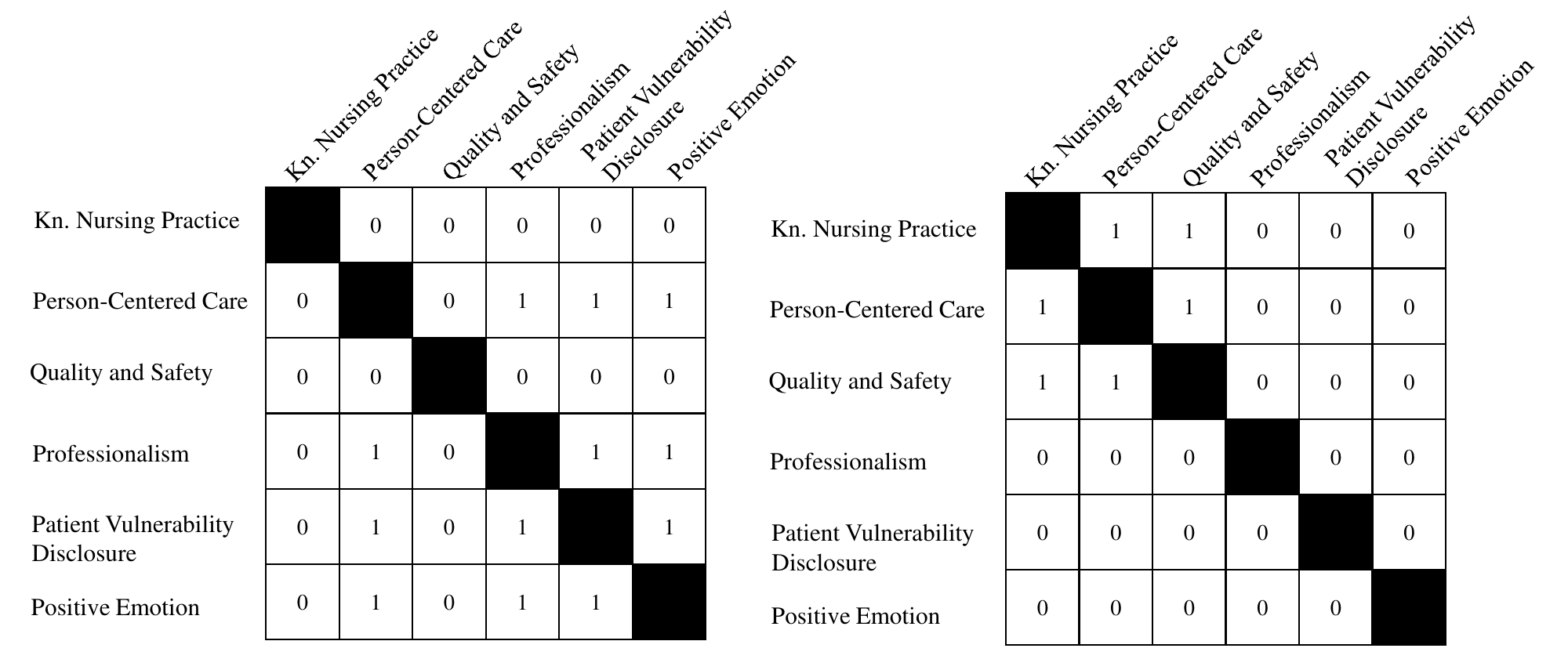}
    \caption{Adjacency matrices, displaying the co-occurrence of codes, for the two stanzas presented in Table \ref{fig:Coded_Data}. The
adjacency matrix has zero values in diagonal cells, and is symmetric.
 }

    \label{fig:Adjacency matrices}
    
\end{figure*}

In terms of data structuring, it is essential to designate one or more columns that define how to divide the data for analysis. These columns are known as stanzas\cite{shaffer2016tutorial}.
Every line within a stanza is related to one another. In other words, elements occurring together within a stanza are conceptually connected. In our analysis, each distinct topic of conversation was treated as a separate stanza.  
Researchers \cite{chesler2015novel, landauer2013handbook, dorogovtsev2003evolution, lund1996producing, sole2001small} have demonstrated that the frequent co-occurrence of concepts within a specific segment is a good indicator of cognitive connections.
Since the ultimate purpose of MENA is to visualize and analyze the structure of connections among elements, finding meaningful connections in the data is critical. In order to achieve these meaningful connections, MENA generates a set of adjacency matrices, each indicating the co-occurrence of codes in a single stanza, (See Fig. \ref{fig:Adjacency matrices}).

That is, the epistemic frame consists of individual frame elements, $f_{i}$, where \textit{i} represents a specific coded element.
For any given utterance \textit{u} within a stanza \textit{s}, each line $L^{u,s}$ provides evidence of the active epistemic frame elements in a specified time window. Once coded, each line of data is represented by a vector of 1s and 0s, indicating the presence or absence of each code, respectively.
These vectors are then transformed into an adjacency matrix $A^{s,p}$ for each stanza \textit{s} made by participant \textit{p}.

\begin{equation}
A_{i,j}^{s,p} = 1 \text{ if } f_i \text{ and } f_j \text{ are both in } L^{u,s}
\end{equation}

In this matrix, if two codes appear together in the same stanza, the cell corresponding to their intersection is set to one. Conversely, cells representing codes that do not appear in the same stanza are assigned a zero.
Since the intersection of codes with themselves does not indicate a relationship between two distinct components, MENA zeroes out the diagonal values \cite{shaffer2016tutorial}. In this way, two stanzas shown in Fig. \ref{fig:Coded_Data}  produce the adjacency matrices shown in Fig. \ref{fig:Adjacency matrices}.

\subsubsection{Accumulation of Adjacency Matrices for Each Unit of Analysis}

To determine the connection structure, MENA aggregates the adjacency matrices of each analysis unit \textit{p} into a single cumulative adjacency matrix, $C_{p}$, where, each cell $C_{p(i,j)}$ indicates the count of stanzas where both codes \textit{i} and \textit{j} appeared: 

\begin{equation}
C_{p} = \sum A^{s,p}
\end{equation}

\begin{figure}[H]
    \centering
    \includegraphics[width=0.4\textwidth]{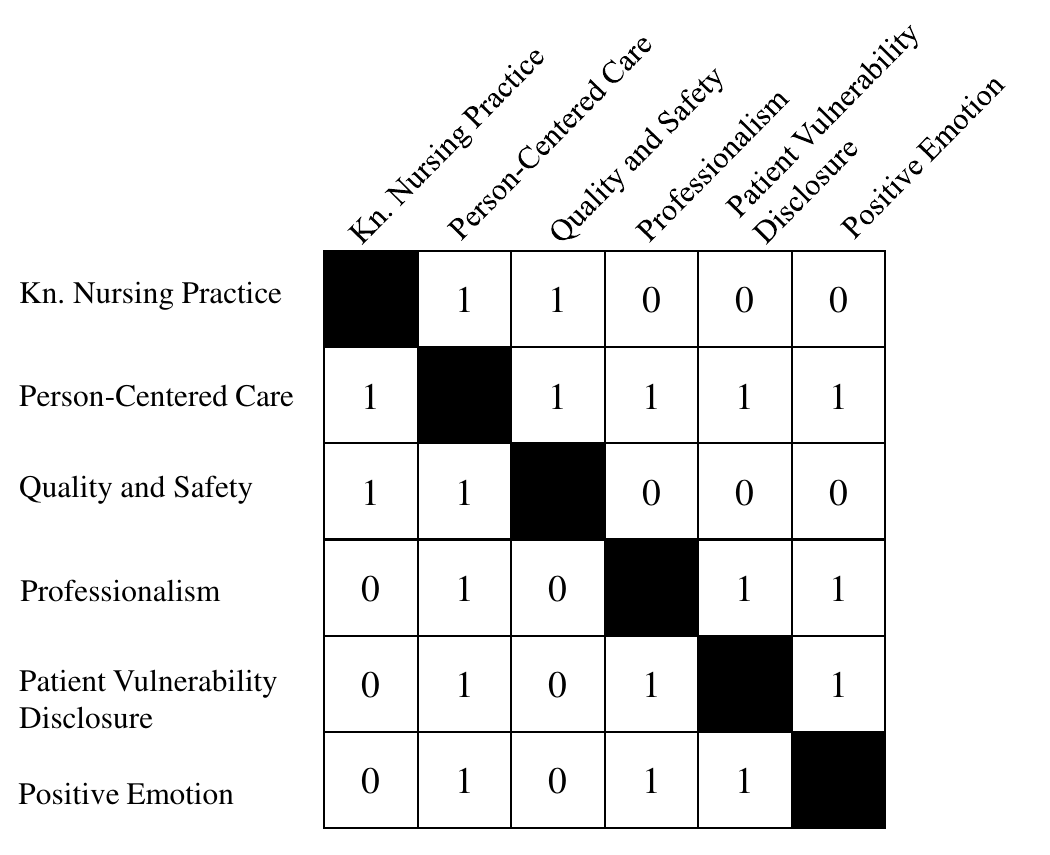}
    \caption{The cumulative adjacency matrix of two stanzas for a participant, identified by Participant1, Which sums the adjacency matrices presented in Fig \ref{fig:Adjacency matrices}.}
    \label{fig:Accumulation Matrix}
    % \Description{This figure displays the cumulative adjacency matrix for the first participant, which sums the adjacency matrices of two stanzas presented in the previous figure. This summation is conducted cellwise across the matrices.}

\end{figure}

In the excerpted dataset shown in Fig. \ref{fig:Coded_Data}, the unit of analysis is each participant across \textit{aware} or \textit{unaware} VGP condition. Since both stanzas shown in Fig \ref{fig:Adjacency matrices} belong to the same participant under the same condition, they are combined as illustrated in Fig. \ref{fig:Accumulation Matrix} by cell-wise addition. So, the total number of matrices to be summed for each analysis unit is equal to the number of stanzas within that unit. Therefore, all matrices from the same condition are aggregated for each participant, ensuring that data from similar conditions are systematically grouped and analyzed together.

Each matrix is then transformed into an adjacency vector, $V_{u}$ by copying the cells from the top diagonal of the matrix row by row into a single vector. It is worth noting that since the cumulative adjacency matrix is symmetric, the vector only includes the top (or bottom) diagonal cells. For instance, the vector $[1,1,0,0,0,1,1,1,1,0,0,0,1,1,1]$ would be used to represent the matrix in Fig \ref{fig:Accumulation Matrix}. These vectors are in a high-dimensional space, \textit{V}, where each dimension denotes a unique pair of two codes. Therefore, the network of connections between objects for each unit is depicted through an adjacency vector in a high-dimensional space, which includes all code co-occurrences accumulated across all stanzas.

\subsubsection{Spherical Normalization of Adjacency Vectors}
In this high-dimensional MENA space, every adjacency vector illustrates the connection pattern of an individual unit, and the length of a vector can be influenced by the total number of stanzas in the analysis unit. To address this issue, MENA spherically normalizes the vectors by dividing each vector $V_{u}$ by its length, ensuring that all vectors have a consistent length:

\begin{equation}
N_u = V_u / \sqrt{\sum (V_u)^2}
\end{equation}

This process normalized the connection strengths to a range of zero to one and accounted for units with different numbers of coded lines. Thus, regardless of the number of stanzas, the normalized vector $N_{u}$ quantifies the relative frequencies of co-occurrence of codes per unit \textit{u}.

This normalized high-dimensional space is called \textit{analytic space}, where each unit of analysis is represented by a single point. Since \textit{analytic space} cannot be directly visualized, a two-dimensional representation of this space is constructed through dimensionality reduction, which will be explained in the following section.

\subsubsection{Dimensionality Reduction via Singular Value Decomposition}

To interpret and visualize the normalized vectors, MENA applies dimensionality reduction to the high-dimensional space by using singular value decomposition (SVD) \cite{bowman2021mathematical}.
SVD produces orthogonal dimensions that maximize the variance in the data explained by each dimension, specifically SVD1 (\textit{X-axis}) and SVD2 (\textit{Y-axis}). As a result, the original high-dimensional MENA \textit{analytic space} is rotated, producing these orthogonal dimensions \cite{shaffer2016tutorial}. For each unit \textit{u} in the dataset, MENA generates a point, $P_{u}$ representing the position of the normalized vector \textit{Nu} within the singular value decomposition framework, known as the \textit{ENA score}.
The resulting points are then visualized by positioning the original frame elements using an optimization routine that minimizes 

\begin{equation}
\sum (P_i - C_i)^2
\end{equation}
where $P_{i}$ is the projection of the point under SVD, and $C_{i}$ is the centroid of the network graph under the node positioning being evaluated.
% These points are utilized to plot individual units of analysis.
In addition, to generate the network representation for all participants within the same condition, MENA sums the cumulative adjacency matrices from each participant, which results in a comprehensive matrix that captures all connections between codes for participants in that condition. The final MENA diagram, derived from this matrix, visually represents the interaction patterns among participants under the same conditions.
Transitioning into the graphical representation, each code from the adjacency matrix is represented as a node in the network \footnote{Fixing the positions of nodes is beyond the scope of this paper; for further details, see \cite{bowman2021mathematical}.}. Links are then constructed between the positioned network nodes according to the adjacency matrix. The size of these nodes (codes) indicates their frequency, illustrating how often they occur within the analyzed data. Meanwhile, the strength of the edges between two codes represents the intensity of their relationship, providing a clear visual indication of how strongly these codes are connected.

\section{Results}

\subsection{RQ1: Assessing the proposed framework for classifying the emotional state }

We pre-trained models on single modalities, including audio, text, and 3D human pose, using datasets such as RAVDESS \cite{livingstone2018ryerson} and CMU-MOSEI \cite{zadeh2018multi}. Subsequently, we integrated these models into a multimodal framework and fine-tuned them with VGP interaction data. Audio data from open-access datasets were converted into MFCC features to train an audio extraction model. For pre-training a pose estimation framework, we trained a human pose estimation model on augmented skeleton pose data from VGP interactions, applying rotational augmentations of $5^\circ$ and $10^\circ$ to help the model generalize better across different views of body movements. For text data, we utilized a RoBERTa model integrated with common-sense knowledge from ConceptNet and pre-trained on the CMU-MOSEI dataset. We employed ConceptNet Numberbatch embeddings for tokens present in the corpus and supplemented missing tokens with FastText embeddings. 

The results in Table \ref{tab:results:combined:classification} show that integrating the audio extraction model, human pose estimation model, and text features (enhanced with the knowledge graph) into a multimodal framework improved the accuracy of the model to 95.76\%.

%The knowledge graph helps in learning the semantics and reasoning beyond text embedding alone. It also helps provide the structural relationship between words and concepts. This common-sense knowledge helps the model understand deeper relationships between words, enriching the comprehension of utterances. 

%The significance of multimodal framework can be observed by results in Table \ref{tab:results:combined:classification}. By combining audio feature extraction, human pose estimation, and RoBerta model with a commonsense knowledge graph, we observed a significant improvement in accuracy for predicting emotional states compared to a model trained on a single modality. This multimodal integration of data modalities provides a deeper understanding of emotional states, as it uses diverse sources of information to capture the nuances of emotions more effectively. 

\begin{table*}[h!]
\centering
\caption{Performance of different data-model combinations for Emotional State Classifier, highlighting enhanced results of the proposed multimodal framework. %over single-modality models.
}
\small % Reduces the text size further
\setlength{\tabcolsep}{10pt} % Adjusts column spacing
\renewcommand{\arraystretch}{1.1} % Adjusts row spacing
\begin{tabular}{llccc}
\toprule
\textbf{S.No} & \textbf{Data-Model Combination} &   \textbf{Data-Modality} & \textbf{Accuracy} & \textbf{F1 Score} \\
\midrule
1  & \textbf{RoBERTa}     &  Text    & 0.7354   & 0.7147    \\
2  & \textbf{CNN-LSTM}     &  Audio      & 0.6974   & 0.6897    \\
3  & \textbf{LSTM}      &  3D Skeleton Coordinates     & 0.7100  & 0.7098    \\
4  & \textbf{RoBERTa+CNN-LSTM+LSTM} &Text-Audio-3D Skeleton Coordinates & 0.9273   & 0.9278    \\
5  & \textbf{RoBERTa+KG+CNN-LSTM+LSTM}   &  Text-Audio-3D Skeleton Coordinates   &  \textbf{0.9576}   & \textbf{0.9550}    \\
\bottomrule
\end{tabular}

\label{tab:results:combined:classification}
\end{table*}

\subsection{RQ2: The structure of connections between nursing competencies of participants and their emotional states}
After applying MENA and setting fixed positions for nodes, we can directly compare different networks in a two-dimensional space by analyzing the connections between nodes and the strength of these connections. Within the network model, the size of a node reflects the relative frequency of the corresponding code, while the thickness and saturation of the lines represent the strength of the connections between two nodes. A thicker line represents a stronger connection while a thinner edge depicts a weaker connection. 

\begin{figure*} 
    \centering     \includegraphics[width=1\textwidth]{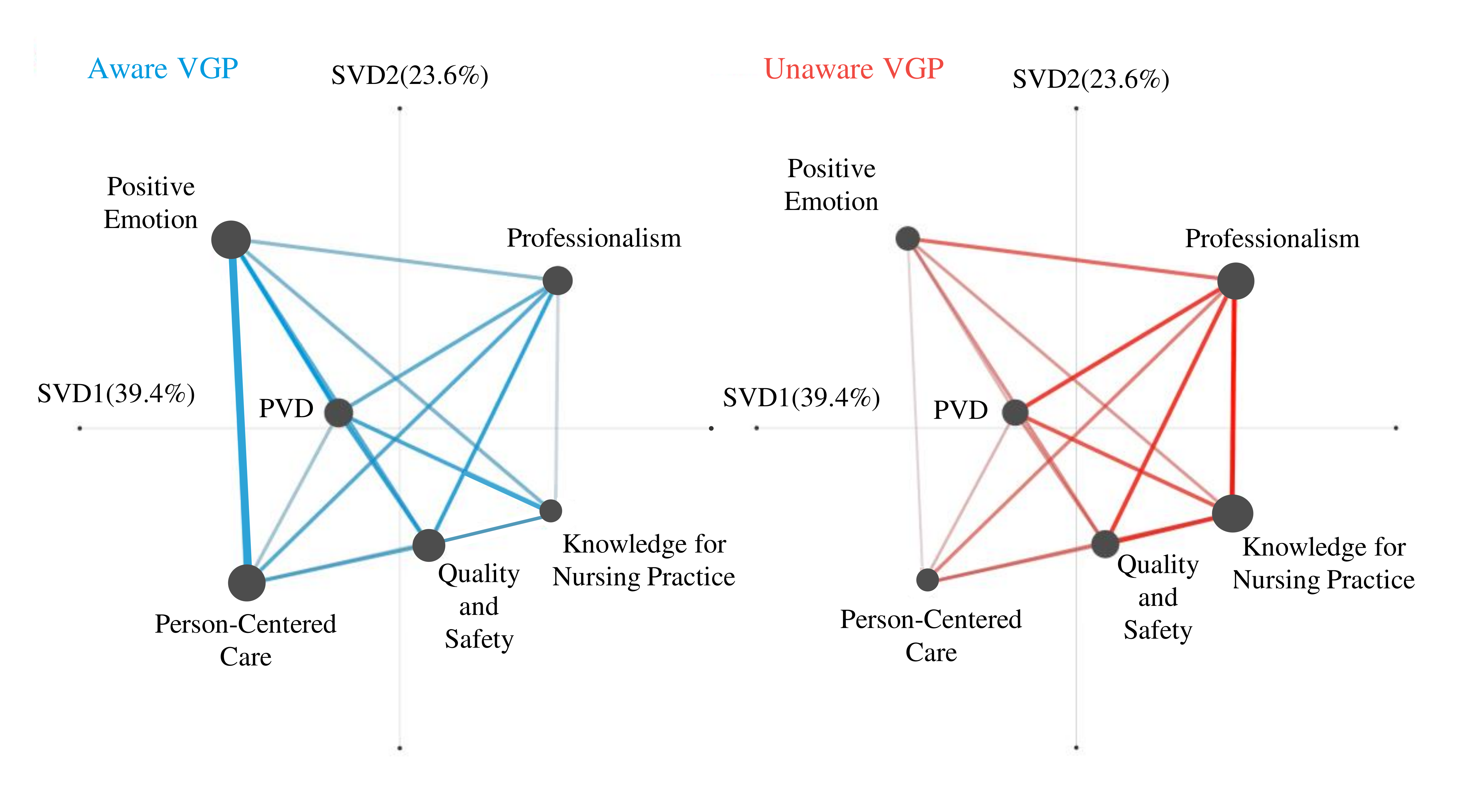}
    \caption{MENA representations of participant behavioral dynamics during interactions with \textit{unaware} (red) and \textit{aware} (blue) VGP. PVD stands for \textit{Patient Vulnerability Disclosure}. In the model, the size of each node corresponds to the frequency with which the code appears, while the thickness of the lines between nodes indicates the strength of the connections. In general, participants demonstrated denser connections on the left side of the \textit{X-axis} in the \textit{aware} condition, exhibiting more positive emotions around \textit{Person-Centered Care} and in response to \textit{Patient Vulnerability Disclosure}. Conversely, in the \textit{unaware} condition, participants showed stronger connections on the right side of the \textit{X-axis}, relying on established protocols.
 }

    \label{fig:MENA Networks}
    
\end{figure*}

Fig \ref{fig:MENA Networks} displays visual representations of the MENA, which show connections among nursing competencies and emotions across participants for each condition. The amount of variance explained by the variables represented in axes SVD1 and SVD2 is 63\% of the total variance: 39.4\% for SVD 1 and 23.6\% for SVD2. Along the \textit{X-axis}, the MENA space is distinguished by more empathetic care standards to the left, e.g., \textit{Person-Centered Care}, and \textit{Positive Emotion}, and more structured professional skills on the right, e.g., \textit{Professionalism} and \textit{Knowledge for Nursing Practice}.
The visualized networks show distinct patterns in the diagrams of each condition. In general, a denser network of connections is evident on the left side of the \textit{X-axis} in the \textit{aware} condition, and participants demonstrated stronger connections between \textit{Positive Emotion} with \textit{Person-Centered Care} and \textit{Patient Vulnerability Disclosure}.
%This suggests that awareness of VGP may influence how participants perceive and engage in caregiving, fostering a more empathetic and person-centered approach.
Conversely, in the \textit{unaware} condition, the strongest connections are between \textit{Professionalism} and \textit{Knowledge for Nursing Practice} with other nodes on the right.
%This suggests that caregivers may rely heavily on their formal training and established protocols in the absence of VGP awareness.

\subsection{RQ3: Differences in the structure of connections when engaging with \textit{aware} vs. \textit{unaware} VGP}

\begin{figure*}
    %\hspace*{-2cm}
    \includegraphics[width=0.9\textwidth]{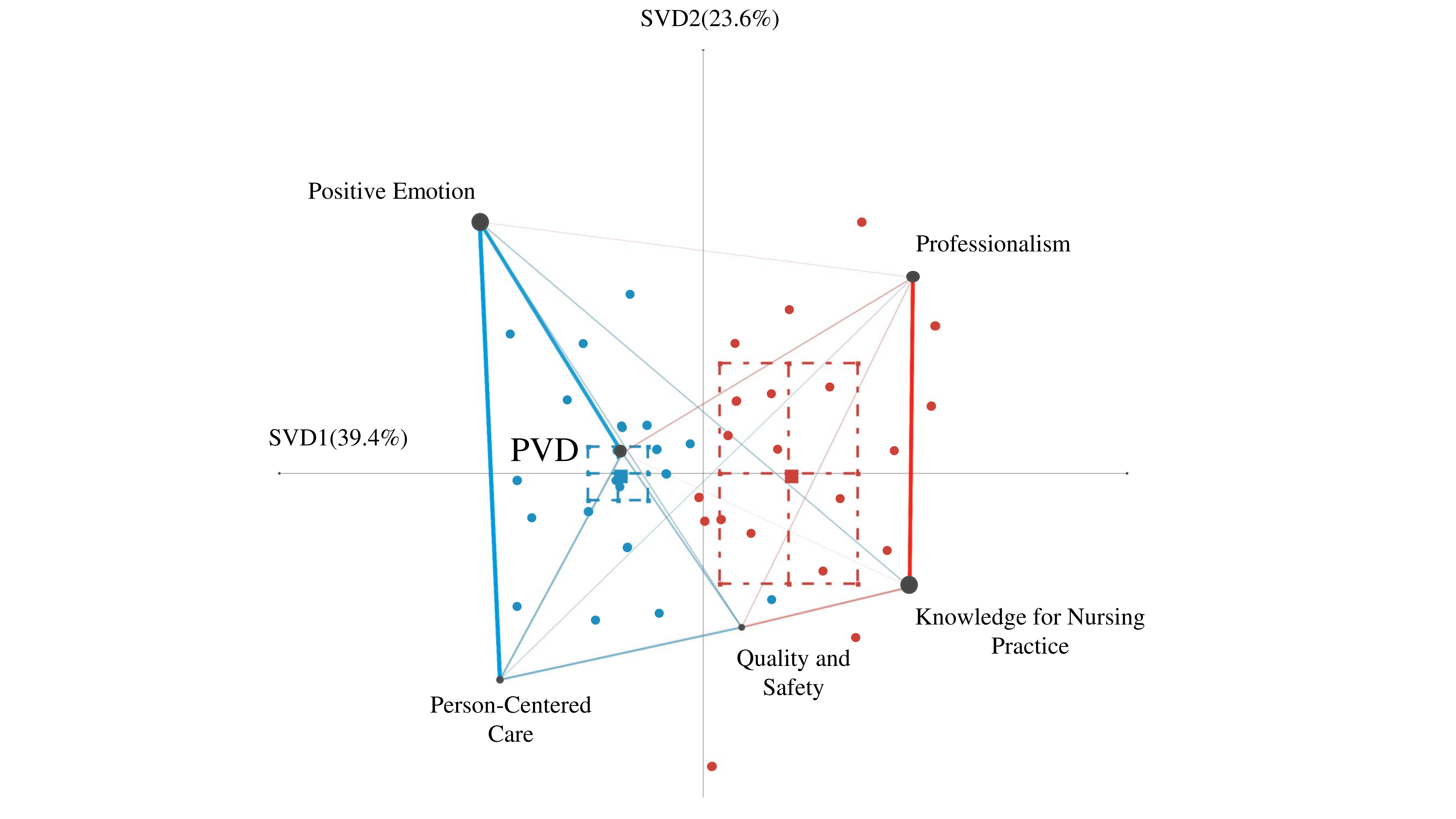}
    \caption{Subtracted MENA graph showing differences between two conditions: \textit{unaware} in red, and \textit{aware} in blue. Line thickness and color saturation represent the relative differences in connection frequencies—red lines for higher frequencies in the \textit{unaware} condition, and blue lines for higher frequencies in the \textit{aware} condition. The points indicate the network positions (\textit{ENA scores}) of participants, with red for \textit{unaware} and blue for \textit{aware} conditions. Colored squares denote the average network positions (Mean \textit{ENA scores}) for each condition, while dashed lines around the means represent the $95\%$ confidence intervals for each dimension.
 }

    \label{fig:Subtracted Networks}
    
\end{figure*}

Fig. \ref{fig:Subtracted Networks} presents subtracted graph that highlights the differences between the two networks along with the \textit{ENA score} for each unit of analysis, which in this case is each participant. This graph is generated by subtracting the weights of each connection in one network from the corresponding connections in the other network. This method reveals which connections are relatively stronger in one network compared to the other one. In this network, centroids are marked with squares, and 95\% confidence intervals are represented by dotted lines.
This figure clearly illustrates the differences in caregiver behavioral dynamics with the VGP under conditions of \textit{aware}, shown in blue, versus \textit{unawar}, depicted in red.
As illustrated in the plot, participants generally formed stronger connections among the constructs in the \textit{aware} mode.

A two-sample t-test was performed to assess if there were significant differences between the means of each condition.
Along the \textit{X-axis}, the results showed a significant difference ($t(55.83) = 2.93, p < 0.01, Cohen’s d = 0.93$) between the \textit{aware} condition ($mean = 0.82 \pm 0.21, N = 20$) and the \textit{unaware} condition ($mean = 0.73 \pm 0.19, N = 20$). This difference was statistically significant at the alpha level of $0.05$. No significant difference was observed along the \textit{Y-axis}.

\section{Discussion and Implications}

The present study introduced Multimodal Epistemic Network Analysis as a new method to explore behavioral dynamics and emotional states of caregivers interacting with a virtual geriatric patient in an AR setting. In addition, it investigated how the awareness of VGP affects caregiver behavior.
By integrating various data modes, each providing unique insights, this multimodal approach enhanced understanding of the interdependence among different types of data, making it more effective than relying on single-mode data.

% By integrating and analyzing multimodal data inputs, MENA provides a more comprehensive understanding of attitudes and emotions compared to traditional ENA. This approach utilizes network analysis to visualize and statistically measure relationships and patterns among various behavioral responses. By understanding how caregivers adjust their behavior in response to different configurations of experiments, training simulations can be specifically designed to enhance caregiving skills.

\textbf{RQ1} focused on developing an Emotional State Classifier to detect the positive emotions of participants. We questioned whether integrating multiple data modalities would improve the performance of our proposed framework. Our findings revealed significant improvements in model integration, suggesting that emotions are associated with various data inputs like audio, 3d human pose, and speech.
Further integrating the multimodal framework with a common-sense knowledge graph resulted in the highest accuracy due to its ability to understand the semantics and reasoning beyond text embedding alone. The knowledge graph also provides the structural relationship between words and concepts, enabling the model to understand deeper relationships between words and thereby enriching its comprehension of utterances. 
%The significance of multimodal framework can be observed by results in Table \ref{tab:results:combined:classification}. By combining audio feature extraction, human pose estimation, and RoBerta model with a commonsense knowledge graph, we observed a significant improvement in accuracy for predicting emotional states compared to a model trained on a single modality. This multimodal integration of data modalities provides a deeper understanding of emotional states, as it uses diverse sources of information to capture the nuances of emotions more effectively. 
Future studies might explore additional factors, such as individual personality traits and environmental influences. These factors could influence how emotions are expressed and interpreted across various populations and settings.

\textbf{ RQ2} asked about the structure of connections among nursing competencies and the emotional state of participants during the study. The prominent link between \textit{Positive Emotion} and \textit{Patient Vulnerability Disclosure} in \textit{aware} mode suggests that participants react to patient concerns with a more supportive and optimistic attitude. This is paralleled by the strong connection between \textit{Positive Emotion} and \textit{Person-Centered Care}, which illustrates that participants are inclined to promote a positive emotional environment around the competency domain of \textit{Person-Centered Care}. Conversely, in \textit{unaware} mode, the strong link between \textit{Knowledge for Nursing Practice} and \textit{Professionalism} suggests that caregivers may rely on a more traditional, possibly rigid, approach to caregiving.
These findings can influence the design of systems that promote more empathetic and attentive care practices, which has potential implications for professional training, especially in fields that demand high levels of interpersonal skills and emotional intelligence.

\textbf{ RQ3} investigated how associations between nursing constructs and emotions vary across two conditions. Employing MENA subtracted network, we observed that participants provided more positive emotional support during their caregiving interactions when the VGP was in \textit{aware} mode. This condition promoted person-centered and empathetic interactions. In contrast, interactions with \textit{ unaware} VGP were more structured and rule-based, focusing on procedural knowledge and adherence to established caregiving protocols.
Thus, the awareness of VGP encouraged caregivers to integrate emotional intelligence and adaptive responses more effectively into their practices. This analysis could guide the development of training strategies that better integrate professional knowledge with adaptive, supportive caregiving techniques. Research findings have highlighted a significant gap in empathy within traditional training methods for caregivers \cite{kiafar2024analyzing}. Thus, incorporating an awareness feature presents a promising opportunity to enhance caregiver preparedness, addressing this deficiency.
%MENA not only demonstrated stronger and more evenly distributed connections in the \textit{aware} condition but also revealed statistically significant differences between the two networks. This approach uncovered differences that were not detectable using the conventional ENA model. Further details about the ablation study are available in Appendix \ref{Appendix1}. Therefore, the proposed framework could be beneficial for analyzing complex interactions in various contexts, providing deeper insights. Its ability to detect subtle differences that traditional models miss suggests
%MENA could be applicable in any environment that relies on dynamic interpersonal interactions such as education and learning.

\paragraph{Limitations and Future Work}
This work is one of the early efforts to use multimodal data analysis within a network-based framework to visualize and present contrastive human behaviors. Like any preliminary work, our work had some limitations. First, the participant pool was predominantly female, reflecting the common gender distribution in this profession. In addition, we focused primarily on the emotional state as an additional modality to enhance the power of the model. In the future, we intend to expand our analysis by incorporating other modalities, such as gaze tracking. These additions can provide a more comprehensive representation of the participants by further enriching the capability of our model.
\section{Conclusion}
In this paper, we introduced a multimodal framework that captures low-level human behaviors and maps them onto high-level visualization diagrams. This framework includes a multimodal emotion classifier component, which enhances the understanding of behavioral dynamics and the interconnection of different constructs. By integrating multiple data sources and analytical methods, the framework offers a detailed view of the complex relationships within human behavior. Applying MENA to interactions of 20 caregivers with a VGP revealed differences that were not detected by traditional epistemic network models. As shown in our visualization results, participants tended to adopt a rigid caregiving approach with an \textit{unaware} VGP and a more emotionally supportive, balanced approach with an \textit{aware} VGP. These insights can inform the development of future geriatric care simulations and tools. Given the analytical and visual flexibility of MENA, it can be applied not only to model processes within virtual settings but also broadly to other application areas and user studies where representation and visualization of meaningful patterns of data associations are demanded for high-level decision-making.

\bibliographystyle{abbrv-doi-hyperref}

\bibliography{template}

\appendix % You can use the `hideappendix` class option to skip everything after \appendix

\end{document}